\documentclass[3p, review, 11pt]{elsarticle}

\usepackage{flushend}
\usepackage{epsfig}
\usepackage{amsmath}
\usepackage{multirow}
\usepackage{amsthm}
\usepackage{amssymb }
\usepackage{amsmath}
\usepackage{bm}
\usepackage{float}
\usepackage{mathrsfs}
\usepackage{amsfonts}
\usepackage{graphicx}
\usepackage{algorithm, algpseudocode}
\usepackage{subfigure}
\usepackage{comment}
\usepackage{verbatim}
\usepackage{xcolor}
\usepackage{rotating}
\usepackage{hyperref}

\def\BibTeX{{\rm B\kern-.05em{\sc i\kern-.025em b}\kern-.08em
    T\kern-.1667em\lower.7ex\hbox{E}\kern-.125emX}}
\def\SMP{\text{COVID19-HPSMP}} 
\def\CDATA{\text{COVID19 PRIMO}}
\def\pt{_{t-1}}
\def\t{_{t}}
\def\x{\bm{x}}
\def\X{\bm{X}}
\def\U{\bm{U}}
\def\e{\bm{e}}
\def\b{\bm{b}}
\def\W{\bm{W}}
\def\NW{N_{W}}
\def\NF{N_{F}}
\def\i{_{i}}
\def\NT{N_{T}}
\def\mT{\mathcal{T}}
\def\l{^{(l)}}

\def\bi{\bm{i}}
\def\g{\bm{g}}
\def\f{\bm{f}}
\def\c{\bm{c}}
\def\h{\bm{h}}
\def\o{\bm{o}}

\journal{Expert Systems with Applications}

\begin{document}

\begin{frontmatter}

\title{$\SMP$: COVID-19 Adopted Hybrid and Parallel Deep Information Fusion Framework for Stock Price Movement Prediction \\
}
\author{Farnoush Ronaghi$\dag$, Mohammad Salimibeni$\dag$,  Farnoosh Naderkhani$\dag$, and Arash Mohammadi$\dag$}
\address{$\dag$Concordia Institute for Information Systems Engineering,  Concordia University,\\
Emails:' $\{$f\_ronagh, m\_alimib$\}$@encs.concordia.ca; $\{$farnoosh.naderkhani, arash.mohammadi$\}$@concordia.ca\\
Coresponding Author: Arash Moohammadi; Email: arash.mohammadi@concordia.ca; Tel: (+1) 514-848-2424 Ext. 2712; Address: 1455 De Maisonneuve Blv. W., EV-009.187, Montreal, QC, Canada, H3G-1M8.}

\begin{abstract}
The novel of coronavirus (COVID-19) has suddenly and abruptly changed the world as we knew at the start of the 3$^{\text{rd}}$ decade of the 21$^{\text{st}}$ century. Particularly,  COVID-19 pandemic has negatively affected financial econometrics and stock markets across the globe. Artificial Intelligence (AI) and Machine Learning (ML)-based prediction models, especially Deep Neural Network (DNN) architectures, have the potential to act as a key enabling factor to reduce the adverse effects of the COVID-19 pandemic and future possible ones on financial markets. In this regard, first, a unique COVID-19 related PRIce MOvement prediction ($\CDATA$) dataset is introduced in this paper, which incorporates effects of  social media trends related to COVID-19 on stock market price movements. Afterwards, a novel hybrid and parallel DNN-based framework is proposed that integrates different and diversified learning architectures. Referred to as the COVID-19 adopted Hybrid and Parallel deep  fusion framework for Stock price Movement Prediction ($\SMP$), innovative fusion strategies are used to combine scattered social media news related to COVID-19 with historical mark data. The proposed $\SMP$ consists of two parallel paths (hence hybrid), one based on Convolutional Neural Network (CNN) with Local/Global Attention modules, and one integrated CNN and Bi-directional Long Short term Memory (BLSTM) path.  The two parallel paths are followed by a multilayer fusion layer acting as a fusion centre that combines localized features. Performance evaluations are performed based on the introduced $\CDATA$ dataset illustrating superior performance of the proposed framework.
\end{abstract}
\begin{keyword}
 COVID-19 Pandemic \sep Deep Neural Networks \sep Hybrid Models \sep Information Fusion \sep Stock Movement Prediction.
\end{keyword}

\end{frontmatter}
\section{Introduction} \label{sec:Introduction}
The novel of coronavirus (COVID-19) has suddenly and abruptly changed the world as we knew at the end of the 2$^{\text{nd}}$ decade of the 21$^{\text{st}}$ century. The global COVID-19 pandemic caused market volatility~\citep{Mazur:2020, Baek:2020} rocketing upward around the world. In particular, the pandemic has negatively triggered several sectors including but not limited to stock markets, global supply chains, labor markets, and consumption behaviors. Disruptions of such sectors, especially the stock markets~\citep{Bustos:2020, Al-Awadhi:2020, Ahmar:2020},   can adversely affect the global economy. The United States volatility levels in the mid-March of $2020$ are similar to those last seen during October $1987$; after $1929$ to $1939$, and;  during in $2008$. In September $2008$, the Dow Jones Industrial Average fell $777.68$ points in intraday trading. During the recent pandemic, in the latter part of March $2020$, volatility began to retreat and, by late April, fell sharply but remained well above pre-pandemic levels. It is expected that the emerging markets, the ones with restricted resources to cope with negative impacts of the COVID-19,  more substantially feel the COVID-19 pressure due to having slower economic growth and not having sufficient capital inflows. In these sad and unfortunate pandemic times, Artificial Intelligence (AI) and Machine Learning (ML)-based~\citep{Radojic:2020, Rezaei:2020, Hoseinzade:2019, Chong:2019, Seong:2021, Zhang:2021} stock market movement prediction solutions can potentially prevent the pandemic crisis that negatively affecting the market across the world causing unexpected havocs.

\vspace{.1in}
\noindent
\textbf{Literature Review:} Stock market movement prediction is a key and challenging problem in financial econometrics as such has attracted extensive recent research focus~\citep{Frankel:1995, Farnoush:2020, Arash:2017, Edwards:2007, Bollen:2011, Jiang:2020, Hu:2019,  Koshiyama: 2020, Schumaker:2009}.  It is widely acknowledged that investors need high-quality data to make informed and accurate decisions. Particularly, in times of market crisis, specifically during the recent COVID-19 pandemic, investors need advanced Big-Data Analytic and Information Technologies~\citep{Choudrie:2021} to acquire timely and accurate data. Using high-quality data, investors  can perform fast analysis and decision making in the market volatility and  react quickly to the fast changing conditions. Any positive or negative news related to the stock market crisis can have a ripple effect on the investors' decision-making process within the stock markets. During the pandemic area, typically, stock movement prediction becomes significantly challenging as stock markets tend to face high fluctuations. Consequently, it is of paramount importance to develop innovative and advanced processing and learning solutions to accurately predict stock movements for achieving maximum potential profit. This has resulted in a recent surge of interest in ML/AI-based prediction techniques~\citep{Hu:2019, Mekayel:2020}  and fusion of multi-modal information sources. In the context of stock price movement prediction, historical stock prices are typically fused with information obtained from media news. For the latter, in addition to the  conventional news platforms, recently, extensive interest is shown towards  utilization of Internet-based news resources, such as social media for development of ML/AI predictive models. The manuscript focuses on this topic and examines the role of COVID-19 related social media news on behavior of Dow Jones market.

Recent advancements and developments in the field of ML and AI, in particular, Deep Neural Networks (DNNs), have motivated different research works to incorporate such advanced modeling techniques for prediction and forecasting tasks in stock markets~\citep{Tetlock:2007}. In particular, there has been a recent surge of interest in information fusion~\citep{Narkhede:2021} and sentiment analysis~\citep{Choudrie:2021} based on stock market data~\citep{Patel:2015, Gite:2021} and their effects on stock markets. On the one hand, Reference~\citep{Patel:2015} focused on predicting direction of movement of stock and stock price index. This study compared different ML models including Artificial Neural Network (ANN), Support Vector Machine (SVM), Random Forests (RF), and naive-Bayes. Two different approaches are developed to feed ML models. The first approach involves computation of ten technical parameters using stock trading data, while the second approach focuses on representing these technical parameters as trend deterministic data. Experiments are performed on the Indian stock markets and based on a dataset consisting of 10 years of historical data from 2003 to 2012 of two stocks (Reliance Industries and Infosys Ltd.) and two stock price indices (CNX Nifty and S\&P Bombay Stock Exchange (BSE) Sensex). Experimental results show that  prediction performance improves when technical parameters are represented as trend deterministic data. Similarly, Reference~\citep{Patel:2015} focused on predicting future values of stock market index via a two-stage fusion approach.  Support Vector Regression (SVR) is used in the  first stage, while ANN, RF, or SVR are used in the second stage resulting in $SVR–ANN$, $SVR–RF$ and $SVR–SVR$ fusion prediction models. The prediction performance of these hybrid models is compared with the stand-alone models illustrating the power of hybrid modeling. On the other hand, Reference~\citep{Gite:2021} focused on predicting stock prices from sentiment analysis with use of ML/DL approaches. News headlines have a tremendous effect on the buying and selling patterns as traders easily get influenced by what they read. In this work, News headlines are extracted from Pulse and historical prices are obtained from Yahoo Finance. The proposed DL models, i.e., LSTM and LSTM-CNN model, performed reasonably well illustrating potentials of DL for sentiment analysis.

DNN-based solutions are data-driven techniques that learn the underlying dynamics of the stock price movements through processing of a large amount of data. DNN-based methodologies are, typically, data hungry and will not perform well in the absence of a large and diversified set of data resources. Availability of public news media, Internet-based news channels, and social media can pave the way to better train DNN models and further increase utilization of AI within stock markets. This research field, however, is still in its infancy due to its high dependence on the reliability and quality of the information available through Internet-based news channels and social media resources~\citep{Hu:2019}. Furthermore, such data sources can not be directly used for prediction tasks~\citep{Luss:2015} due to the highly correlated nature of stock market price movements. To tackle the aforementioned issues, there is an unmet and timely quest to  develop and design: (i) Hybrid processing/learning models based on different and diversified learning architectures to capture underlying correlations and variabilities of the data sources, and; (ii) Smart fusion strategies to combine scattered social media news with historical mark data. The main objective of the proposed DNN-based predictive model is to construct a new information fusion framework to analyze and interpret ever-changing trends during the COVID-19 pandemic area.

\vspace{.1in}
\noindent
\textbf{Contributions:}  The paper makes the following contributions. First, a unique and real COVID-19 related PRIce MOvement prediction ($\CDATA$) dataset~\citep{CDATA}\footnote{The $\CDATA$ dataset is accessible  through the following page:\\ https://github.com/MSBeni/COVID19\_PRIMO\#COVID19-PRIMO.}'' is constructed to incorporate effects of internet-based and social media trends related to COVID-19 on stock market price movements.  The main  component of the constructed $\CDATA$ dataset is based on Twitter messages. It is well known that news and media move stock prices~\citep{Fama:1998, Huang:2020, Yun:2019}. Nowadays, information reaches out to the public via different news platforms ranging from newspaper, radio and television to social media and Internet-based venues. In this area, social media, especially Twitter, is a popular and widely used platform to share personalized opinion on different topics. Twitter is also used extensively by politicians who potentially have high impact on stock price movements. Based on a survey on Statista~\citep{Clement:2020}, from the first quarter of 2017 to 2020, Twitter had $186$ million active users worldwide.

Based on the constructed $\CDATA$ dataset, the paper proposes a data-driven (DNN-based) COVID-19 adopted Hybrid and Parallel deep  fusion framework for Stock price Movement Prediction ($\SMP$)  that uses information fusion to combine COVID-19 related Twitter data with extended horizon market historical data. More specifically, in contrary to the existing data-driven movement prediction models, where a single DL model is used~\citep{Farnoush:2020},  the proposed $\SMP$ is a hybrid framework with two parallel paths, i.e., one based on Convolutional Neural Network (CNN) with Local/Global Attention modules, and one integrated CNN and Bi-directional Long Short term Memory (BLSTM). The former path is incorporated within the $\SMP$ framework to extract temporal features, while the latter path  is used to extract spatial features. The two parallel paths are followed by a multilayer fusion layer acting as a fusion centre that combines localized features extracted in each of the two parallel paths. The $\CDATA$ dataset is used to evaluate the performance of the proposed $\SMP$ framework, which illustrates its superior performance compared to its stand-alone (non-hybrid)  counterparts.

The remainder of the paper is organized as follows:  Section~\ref{sec:measure} introduces the $\CDATA$ dataset and formulates the stock movement prediction task. The $\SMP$ hybrid framework is presented in Section~\ref{sec:FingPrnt}. The implementation study and results are presented in Section~\ref{sec:exp}. Finally, Section~\ref{sec:conc} concludes the paper.

\section{Problem Definition and $\CDATA$} \label{sec:measure}
In this section, first, the $\CDATA$ dataset is introduced, which is constructed based on the Dow Jones stock market index and its associated Twitter messages for the period of 01/01/2016 to 30/07/2020. The focus is on the problem of stock price movement prediction as close observation of market movements can reveal presence of a significant amount of trading targets with minor movement ratios. More specifically, the paper focuses on investigating effects of COVID-19 pandemic on stock price movement prediction. In this paper, stock movement prediction is modeled as a two-class classification problem based on the adjusted closing price of the underlying stocks. The adjusted closing price is commonly utilized to compute the associated stock dividends and earnings~\citep{Xie:2013}. Furthermore, the adjusted closing price is beneficial to learn and predict fluctuations in the stock market~\citep{Li:2014, Rekabsaz:2017}.

We have prepared a new dataset for the aforementioned prediction problem, which can facilitate analysis and evaluation of potential impacts of a pandemic on stock market and can provide priceless insights to combat future possible pandemic. The constructed $\CDATA$ dataset consists of two components, i.e., historical prices and Twitter messages. The first component, historical data, is obtained from Dow Jones stock market. With the ticker of DJI.  Dow Jones is a stock market index that measures the performance of $30$ large companies like Apple, Boeing, and Microsoft. Historical stock market prices are obtained from the Yahoo finance. For this task, we used the Yahoo finance library in Python\footnote{http:https://pypi.org/project/yahoo-finance/} to collect the data from the Yahoo API. For some of the stocks, we also used  Alpha Vantage APIs\footnote{http:https://pypi.org/project/alpha-vantage/}. The data is prepared based on  three different temporal resolutions, i.e., daily; weekly, and; monthly. The daily prices are used in our model described later for the task of stock movement prediction.

\begin{figure}[t!]
\centering
\includegraphics[scale=0.4]{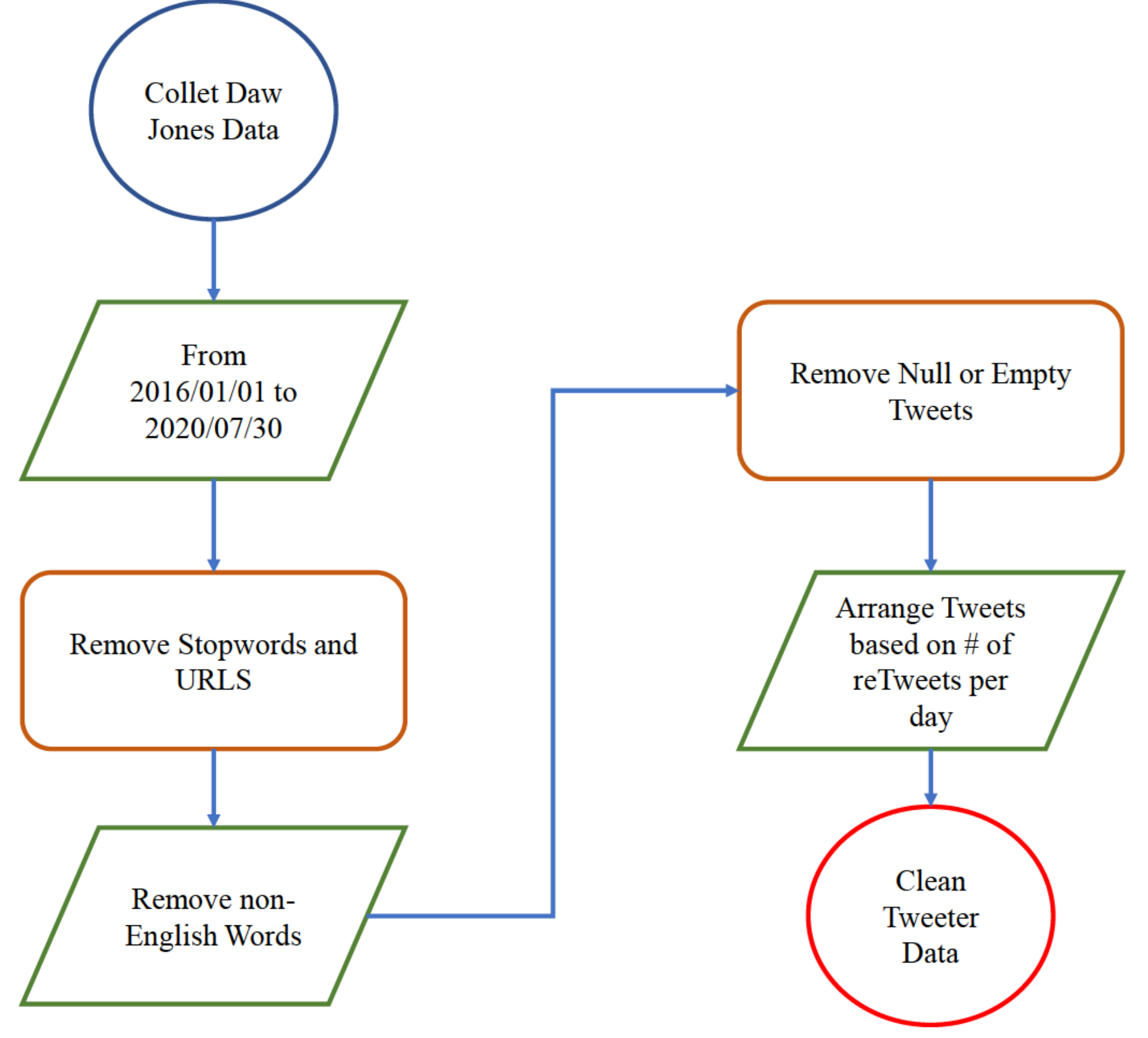}
\caption{ Block diagram of the procedure designed to collect and prepare Tweet component of the $\CDATA$. \label{Fig:NA1}}
\end{figure}
\begin{figure}[t!]
\centering
\includegraphics[scale=0.27]{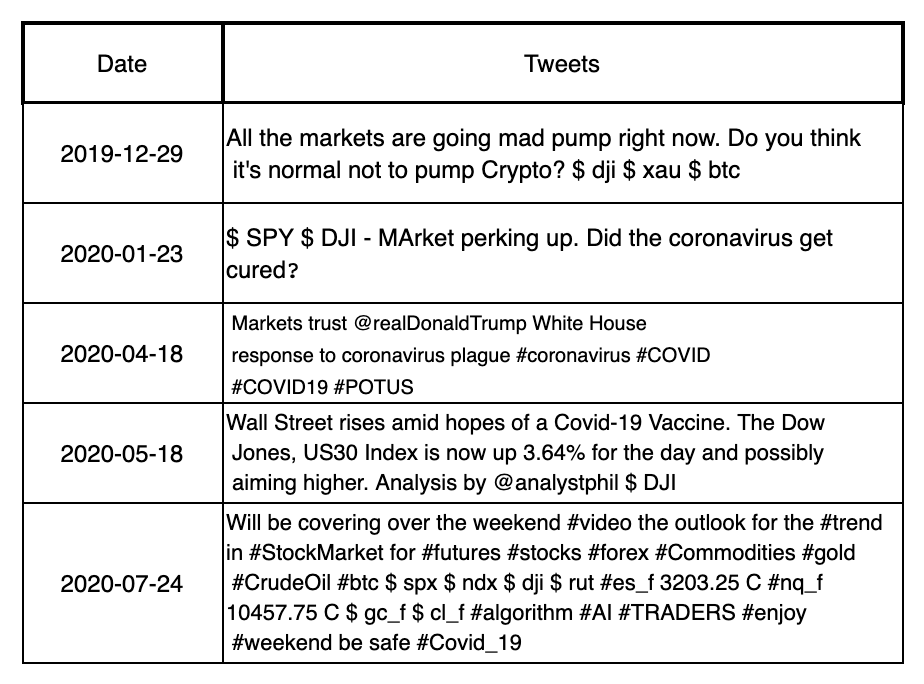}
\caption{Illustrative raw Tweets samples in the $\CDATA$ dataset before the pre-processing step. \label{Fig:NA2}}
\end{figure}

Capitalizing on the facts identified in Section~\ref{sec:Introduction}, for the news component of the COVID-19 price movement prediction dataset, we focused on Twitter. Fig.~\ref{Fig:NA1} shows the block diagram of the approach followed to collect and analyze Twitter messages. Web scraping from the Twitter search engine is utilized to build the Twitter dataset. The official API of the Twitter has some limitations that restricts the extent of text that can be extracted. Additionally,  the official API of the Twitter cuts the tweets at times, which in turn results in items with missing data. We have developed a localized API to address the aforementioned issues. The localized API uses Twitter search engine and directly collects the required dataset from Twitter. We set up our data collection platform based on scraping the twitter website. The twitter web scraping returns the Tweet text content with a range of useful attributes, for example, $Tweet-ID$, Tweet Created at, Retweet, Text, Favorite Count, Hashtag Text, User ID, Followers Count, Friends Count, Statuses Count, User Created at, and Location. To collect informative public Tweets, we added a constraint to our implementation to collect tweets retweeted more than once. Many other unnecessary attributes regarding a tweet were also removed from the data gathering session to focus on the essential information such as date, tweet text, and number of retweets.  Fig.~\ref{Fig:NA2} illustrates an illustrative example of raw tweets collected by web scraping.

\begin{figure}[t!]
\centering
\mbox{\subfigure[]{\includegraphics[scale=0.52]{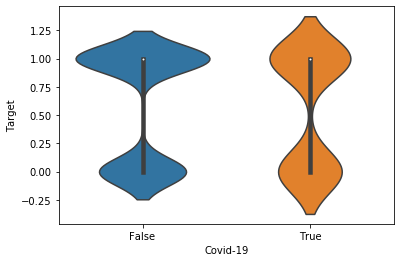}}
\subfigure[]{\includegraphics[scale=0.52]{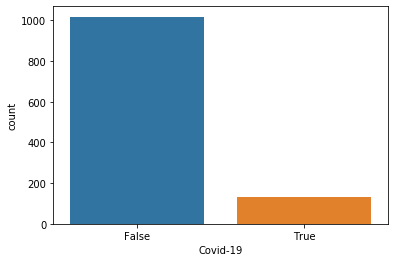}}}\\
\mbox{\subfigure[]{\includegraphics[scale=0.42]{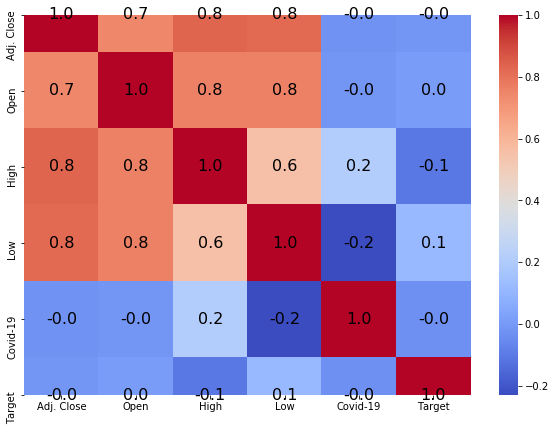}}}
\caption{(a) The Distribution of The COVID-19 Tweets and Target. (b) The Number of  COVID-19 Tweets. (c) Correlation matrix between different variables.}
 \label{Fig:NA3}
\end{figure}
A critical challenge is  scraping the raw content of Twitter data. Such a process takes extensive time and needs manual and cumbersome pre-processing procedures. We retrieved Dow Jones and $COVID-19$ tweets by querying symbols $\$DJI$, $\#DOW$, $Covid19$, $\#Covid19$, $\#Covid-19$ and $CoronaVirus$.  Additionally,  the corresponding data associated with historical prices are collected. The constructed dataset includes related tweets from 01/01/2016 to 30/07/2020 for the Dow Jones stock index. Not every day is considered as a trading day, i.e., weekends and holidays are not among the trading dates and ought to be out of the analysis scope. To better organize and use the input, we subtract the number of days in a year from the number of weekends, the number of half trading days, and the number of market holidays. More specifically,  our dataframe is created by combining historical prices and Tweet corpora and matching them to the trading days. Consequently, we considered $1,152$ trading days from January 2016 to July 2020 to build our dataset. The $\CDATA$ dataset is then divided into a training set from January 2016 to January 2020, and a validation set from January 2020 to February 2020. Data from 01/03/2020 to 30/07/2020  is kept to be used for test purposes.

\subsection{Data Visualization}
\begin{figure}[hp!]
\centering
\includegraphics[scale=0.45]{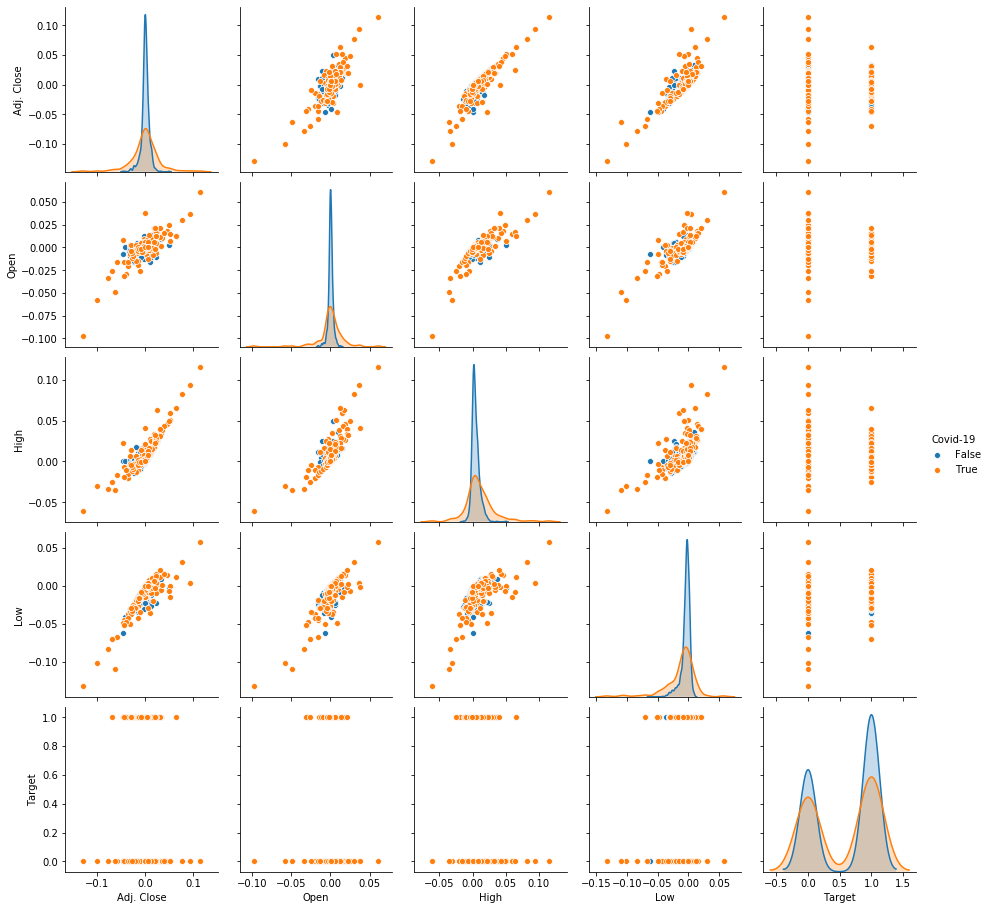}
\caption{Grid of scatter plots.  The orange dots show COVID-19 related data points, while the blue dots represent  lack of COVID-19 related data. \label{Fig:NA4}}
\end{figure}

In this sub-section, we visualize the existing relations between different parameters of the $\CDATA$ dataset, particularly that of COVID-19 tweets with other parameters. As stated above, the introduced dataset is constructed based on the Dow Jones stock market index and its associated Twitter messages for the period of 01/01/2016 to 30/07/2020. The COVID-19 pandemic crisis covers a fraction of the data represented in the $\CDATA$ dataset but plays an essential role in predicting the pandemic's effects on the stock market movements. The COVID-19 related tweets appeared in 2020 (from February to July), specifically starting to show up from the end of February. The dates containing the COVID-19 are stamped with True, making it possible to consider its distribution in the whole frame.  Fig.~\ref{Fig:NA3} visualizes different aspects of the $\CDATA$ dataset and illustrates  relation of COVID-19 with other parameters of the dataset. Violin plot is shown in Fig.~\ref{Fig:NA3}(a), where the distribution of COVID-19 tweets and the market movements are illustrated. As shown in  Fig.~\ref{Fig:NA3}(b), the COVID-19 tweets are less than $200$ days, which is expected given the recent emergence of the pandemic.  Fig.~\ref{Fig:NA3}(c) shows the correlation between several different variables affecting the $\CDATA$ dataset.  Market prices, including Adjusted close price (Adj. Close); Open price (Open); High price (High); Low price (Low); calculated target (Target), and; Presence of COVID-19 in tweets are depicted in this figure. For example, the correlation between normalized adjusted close price and normalized high price is $0.83$.  Finally, Fig.~\ref{Fig:NA4}  is a grid of scatter plot used to visualize bivariate relationships between combinations of variables. Fig.~\ref{Fig:NA4} is included to have a big picture of the distribution of the data and better understand existing relations between different parameters in the dataset. Fig.~\ref{Fig:NA4} shows the relationship for a different combination of variables in a DataFrame as a matrix of plots. The orange dots, show the COVID-19 related data in the dataset, while the blue dots, represent the lack of COVID-19 related data. Fig.~\ref{Fig:NA4} can potentially depict the bivariate relationships between different market price data and COVID-19 together with the relation between the Target recognized in this period of time with the pandemic data.

\section{Proposed $\SMP$ Framework} \label{sec:FingPrnt}
\begin{figure}[th!]
\centering
\includegraphics[scale=0.5]{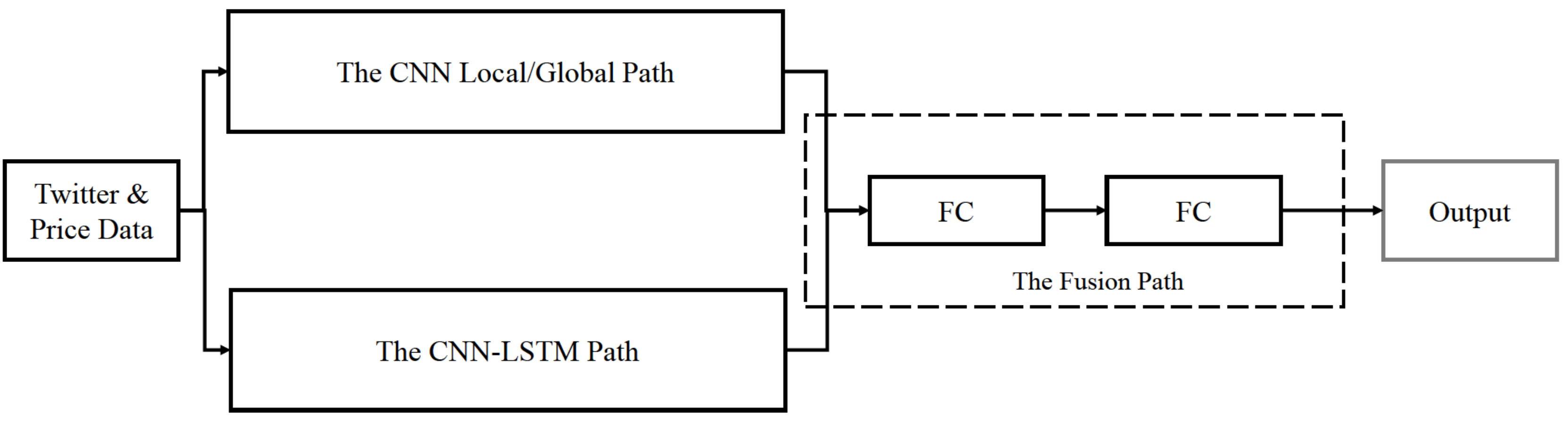}
\caption{The proposed $\SMP$ framework. \label{Fig:1n}}
\end{figure}
In this section, we describe the constituent components used in the development of the proposed $\SMP$ framework. As stated previously, the main architecture of the $\SMP$ is developed based on DNNs. The prominent advantage of DNNs is their ability to extract meaningful patterns from raw data through multiple non-linear transformations and approximation of complex non-linear functions~\citep{Al-Dulaimi:2019}. More specifically, the proposed $\SMP$ is a data-driven (deep learning model) designed based on hybrid or multiple-model strategies. The $\SMP$ framework extracts and interprets the available news corpus via temporal attention modeling based two key principles, i.e., ``Diverse Influence'' and  ``Sequential Context Dependency''. To achieve these objectives, the $\SMP$ is designed as a hybrid multi-modal fusion framework that integrates information obtained from stock market historical data and social media (Twitter data). The proposed hybrid framework consists of three paths, two parallel paths, i.e., the CNN Local/Glocal path, and; the CNN-LSTM path, together with a fusion path. Fig.~\ref{Fig:1n} illustrates the overall structure of the proposed framework. The fusion path composed of fully connected layers that combine extracted features from each of the two parallel paths. Each of the two parallel paths within the $\SMP$ framework are constructed based on the following two main components:
\begin{itemize}
\item[(i)] \textit{Word Embedding Module}: This module is used  to calculate embedded vectors for Twitter data. For this purpose, Glove~\citep{Pennington:2014}, as a pre-trained unsupervised model, is used within the word embedding module of each of the two parallel paths within the proposed $\SMP$ framework.

\item[(ii)] \textit{Attention Module}: The main objective of this module is to extract specific words with highest attention weight. The $\SMP$ is a hybrid model where each of the two parallel paths (i.e., the CNN Local/Glocal path, and; the CNN-LSTM path) is a unique attention module extracting different related features. The rationale behind such a hybrid and parallel structure is the significance of the attention network and the intuition that extracting different attention-related features would improve the overall performance of the model.
\end{itemize}
In what follows, we present each of the three constituent paths of the proposed $\SMP$ framework.

\subsection{The CNN Local/Global Path} \label{sec:LG}
\begin{figure}[t!]
\centering
\includegraphics[scale=0.065]{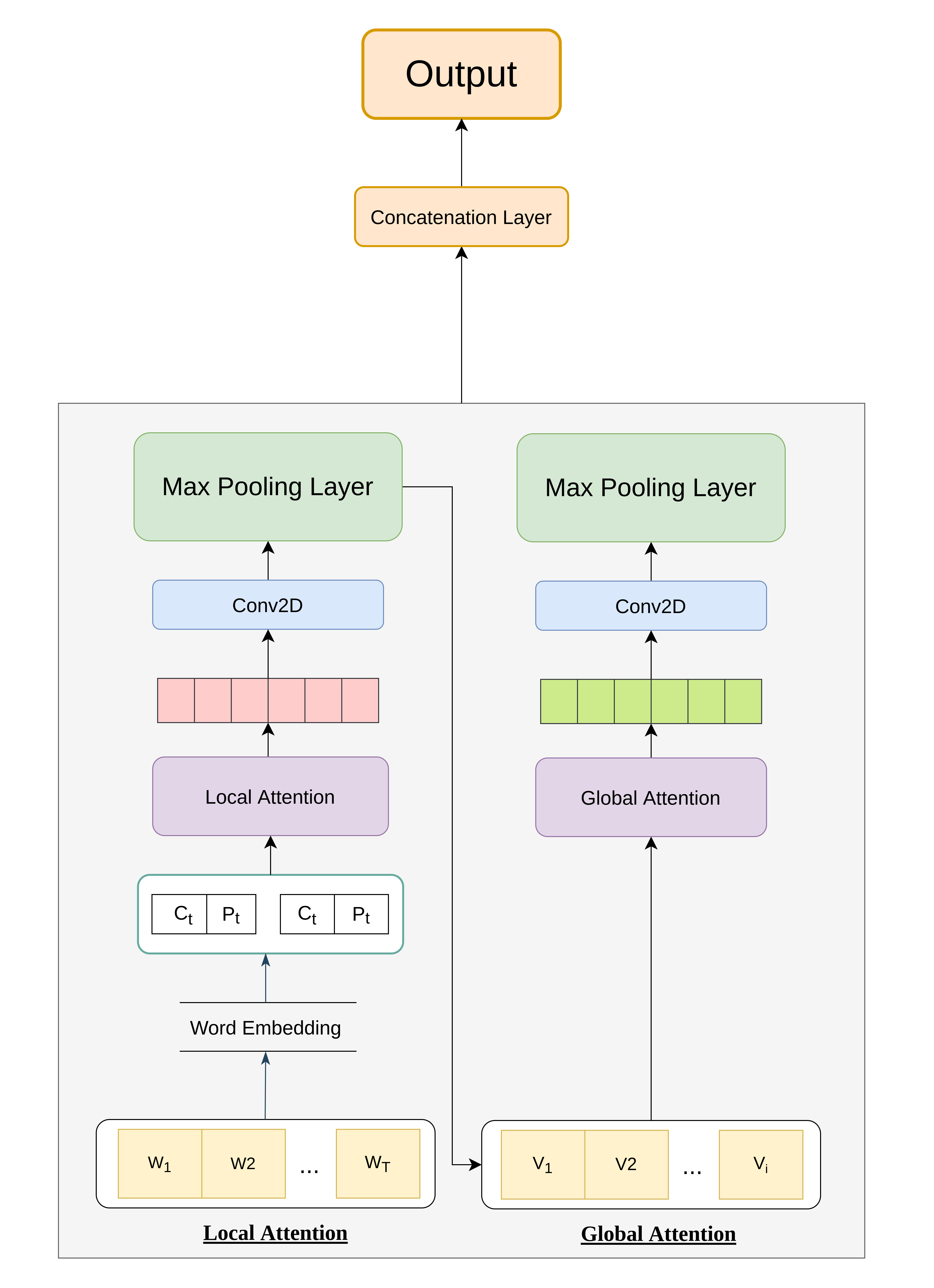}
\caption{The CNN Local/Global path of the $\SMP$ framework. \label{Fig:2n}}
\end{figure}
The first parallel path of the proposed $\SMP$ framework is a CNN-based local/global attention model designed to capture and extract spatial features from the input data. More specifically, the CNN-based path consists of  Local and Global attention layers, which are described in details below.

\subsubsection{Local Attention Layer (LAL)}\label{sec:exp}
Intuitively speaking, a word embedding model produces representations for each word in the Twitter corpus. Let us denote the $l^{\text{th}}$ Tweet among the available set of $\NT$ Tweets with $\mT\l$, for ($1 \leq l \leq \NT$). Furthermore, consider that Tweet $\mT\l$ contains $\NW\l$ number of words. The embedding can be thought of as a linear operator (function) that takes as an input a one-hot vector $\bm{e}\l\i \in \mathbb{R}^{\NW}$ corresponding to the $i^{\text{th}}$ word of Tweet $\mT\l$, for ($1 \leq i \leq \NW\l$). Note that, here $N_W$ denotes the number of words in the overall vocabulary. The embedding then maps the one-hot vector into a dense feature vector $\x\l\i = [x\l_{i,1}, \ldots, x\l_{i,\NF}]^T \in \mathbb{R}^{\NF}$, which consists of $\NF$ scalar features for the $i^{\text{th}}$ word of the $l^{\text{th}}$ Tweet within the Twitter corpus. The feature vector $\x\l\i$ is obtained based on a trainable weighting matrix (to be learned) of the embedding layer as follows
\begin{eqnarray}
\x\l\i = \W^{(Emb)} \e\l\i.
\end{eqnarray}
The embedding layer's output together with price values are provided as a concatenated input to the Local Attention Layer (LAL). The LAL focuses on the words, which are more informative within a localized window.  More specifically, let the $l^{\text{th}}$ Tweet be represented by $\NW\l$ word embedding as ($\x\l_1, \ldots, \x\l_m, \ldots, \x\l_{\NW\l}$), where $\x\l_m$ is the middle (center) word within the embedding sequence of the $l^{\text{th}}$ Tweet. Local attention process is achieved via a sliding window of length $W$ rolling over the word embedding sequence of $\mT\l$. Attention score $s\l\i$ for the $i^{\text{th}}$ word of $\mT\l$ is computed based on an attention weighting score $\W^{(l, LA)}\i \in\mathbb{R}^{W\times \NF}$, for ($1 \leq i \leq \NW\l$), and its associated bias vector $\b^{(l, LA)}\i$ as follows
\begin{eqnarray}
s^{(l, LA)}\i = \sigma\Big(\X^{(l, LA)}\i \circ \W^{(l, LA)}\i + \b^{(l, LA)}\i\Big),  \label{eq:n2}
\end{eqnarray}
where $\circ$ denotes the Hadamard product (element-wise multiplication), $\sigma{\cdot}$ is the sigmoid activation function, and
\begin{eqnarray}\label{Eq:pathloss}
\X^{(l, LA)}\i \triangleq \big[\x\l_{m+\frac{-W+1}{2}}, \ldots \x\l\i, \ldots \x\l_{m+\frac{W+1}{2}}]^T,  \label{eq:n3}
\end{eqnarray}
where superscript $T$ denotes transpose operator. The attention score $s\l\i$ is used as a weight for the words to form localized word embedding as follows $\hat{\x}\i^{(l, LA)} = s^{(l, LA)}\i \x\l\i$.  A higher attention score can be interpreted as higher importance associated with that specific word than the others. The weighted sequences then go through a Convolutional layer with a kernel size of $15$, which is designed to avoid overfitting. A Max-Pooling layer is then implemented after the convolution one to creates position invariance over larger local regions and down-sample the input. Addition of the Max-Pooling layer also leads to a faster convergence rate by selecting superior invariant features, which in turn improves generalization performance.

\subsubsection{Global Attention Layer (GAL)}\label{sec:exp}
The output of the LAL is the provided as input to a Global Attention layer (GAL).  This scoring process of the GAL is similar in nature to that of the LAL (Eqs.~\eqref{eq:n2}-\eqref{eq:n3}). However, the attention score, now denoted by $s^{(l, GA)}\i$, is computed through the entire input, i.e.,
\begin{eqnarray}
s^{(l, GA)}\i = \sigma\Big(\X^{(l, GA)} \circ \W^{(l, GA)}\i + \b^{(l, GA)}\i\Big),
\end{eqnarray}
where $\W^{(l, GA)}\i \in \mathbb{R}^{\NW\l\times \NF}$, and
\begin{eqnarray}\label{Eq:pathloss}
\X^{(G, Att)} \triangleq \big[\x_{1}, \ldots, \x_{\NW\l}\big]^T.
\end{eqnarray}
 By applying global attention, the effect of uninformative words will be diminished, and the global semantic meaning will be captured more precisely through the CNN path. This completes description of the CNN Local/Global path of the proposed $\SMP framework$. Next, we present the CNN-BLSTM path.

\subsection{CNN-LSTM Attention based Model} \label{sec:ME}
\begin{figure}[t!]
\centering
\includegraphics[scale=0.3]{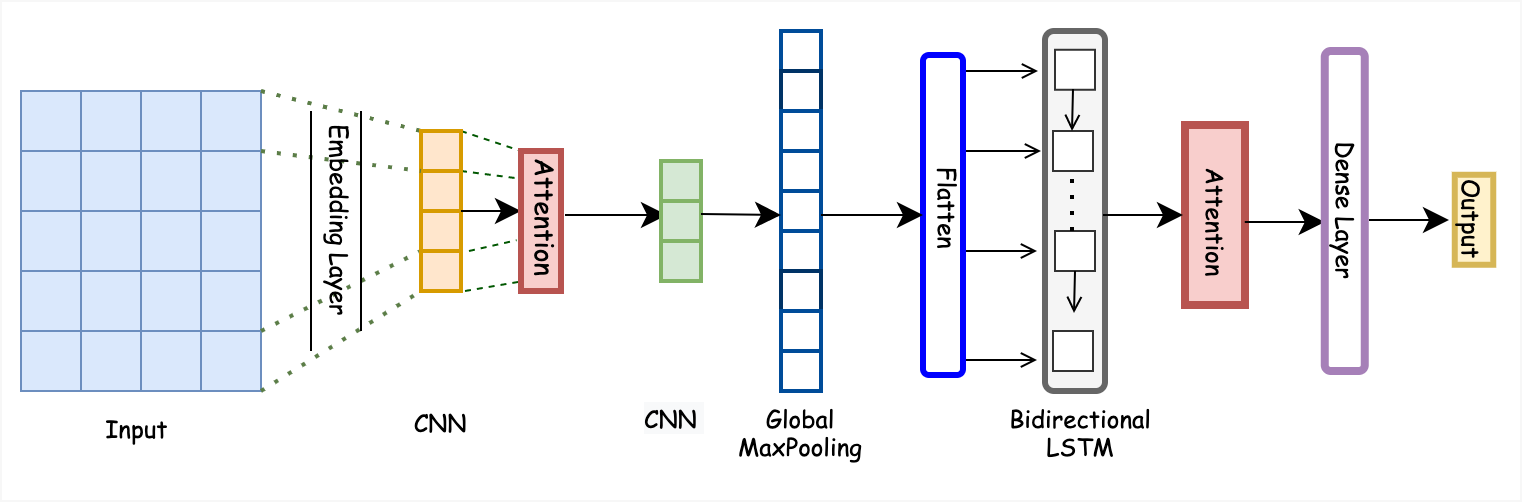}
\caption{The CNN-LSTM path of the $\SMP$ framework. \label{Fig:3n}}
\end{figure}
The second parallel path of the proposed $\SMP$ is a hybrid CNN and BLSTM attention model, referred to as the CNN-BLSTM path. Similar to the CNN Local/Global path, in the first step, the Twitter messages are provided as input to a ``Word Embedding layer". As stated previously, a pre-trained unsupervised Glove model~\citep{Pennington:2014} is used as the word embedding layer within the $\SMP$ framework. Afterwards, the corpus and prices are encoded by a CNN layer to extract general contextual features. An attention layer is assigned across all the vectors to calculate the weighted corpus. At the next step, a second CNN layer is implemented to capture and learn more fine-tuned  features.
The first CNN layer has $50$ number of filters with a window size of $25$. The second CNN layer has $100$ filters with a window size of $25$. The first Attention layer is used to capture  essential and unique features to provide insight into the vector of the data including tweets and prices. The second Attention layer acts on each vector and calculates the weighted mean of these encoded corpus vectors to represent the overall sequential context information. A global Max-Pooling layer is then applied  to capture the essential features and reduce the framework's complexity. Global Max-Pooling is similar to the regular version but with pool size equals to the size of the input. At the next stage, an attention-based BLSTM layer is designed to remember what has been previously learned to better understand the input. The attention-based BLSTM layer is described next.

\subsubsection{Attention-based Bidirectional LSTM:}
To encode temporal information based on the available set of news corpus and financial time-series data, BLSTM is incorporated within the $\SMP$ hybrid framework. Learning based on financial time-series data is a sequence learning task for which BLSTMs are considered as the state-of-the-art DNN architectures.  The LSTM architecture is initially developed by Hochreiter and Schmidhuber~\citep{Hochreiter:1997} to address the vanishing and exploding gradient problem of conventional Recurrent Neural Networks (RNNs). Since then, LSTM models have gained significant popularity owing to their extensions, advancements and successful applications in different domains. Generally speaking, LSTM is a memory-based architecture that uses different gating functions and a memory state to manage process if information through time~\citep{Di:2018}. LSTM works based on the following update model at each time step (denoted by $t$)
\begin{eqnarray}\label{Eq:Bidirectional LSTM}
\bi_t &=& \sigma(\W\i x\t + \U_i \h\pt +\b_i)\\
\f_t &=& \sigma(\W_f x\t + \U_f \h\pt +\b_f)\\
\g_t &=& \tanh(\W_c x\t + \U_c \h\pt +\b_c)\\
\c_t &=& \f_t \circ \c\pt + \bi_t \odot \g_t\\
\o_t &=& \sigma(\W_o x\t + \U_o \h\pt +\b_o)\\
\h_t &=& \o_t \circ \tanh(c_t)
\end{eqnarray}
where $\W\i$, $\W_o$, $\W_f$, $\W_c$, and $\U\i$, $\U_o$, $\U_f$, $\U_c$ are weight matrices; Terms $\b\i$, $\b_o$, $\b_f$, $\b_c$ are bias vectors, and; $\tan(\cdot)$ represents element-wise hyperbolic tangent activation function. Furthermore, $\h\t$ and $\h\pt$ represent the current and previous hidden states, respectively. In the context of the proposed $\SMP$ and to encode the temporal layer, we adopt Bidirectional version of the LSTM (BDLSTM) to feed the $i^{\text{th}}$ word embedding. BLSTM can access both the preceding and succeeding contexts. It separates the hidden layer into two parts, forward state sequence and backward state sequence based on an iterative process.

\subsection{Fusion Path} \label{sec:ME}
The final component of the proposed $\SMP$ framework is the Fusion Path with three fully connected layers for fusing features extracted from each of the two underlying parallel paths and performing the final price movement prediction task. The first fusion layer has $100$ number of neurons and uses ``$\tan$" activation function, while the second fusion layer has  $50$ number of neurons with the same activation function. The final layer of the Fusion Path, has $1$ neuron and uses Rectified Linear Unit (ReLU) as its activation function to produce the price movement predictions. The input to the Fusion Path is constructed by concatenating the output of the CNN Local/Global path, which is a flattened 1-Dimensional feature vector, with that of the CNN-BLSTM path.

As a final note, we would like to briefly  focus on trading from financial perspective to better link the $\CDATA$ (presented in Section~\ref{sec:measure}) and the $\SMP$ framework (discussed in the section). The $\CDATA$ provides means to trade based on  5-day in-advance predictions (provided by the $\SMP$). Generally speaking, a common reason for purchasing, conventionally used by traders and stockholders, is to only buy a stock when its prediction is increasing as this provides the chance to sell at a higher price in future and gain profit. This future to sell a stock, varies for different investors, e.g., some consider a short period trade, some are more eager to mid-term trades, and others  focus on long-term trades. A sentimental engine that can infer the market's sense and predict its movements is a significant asset. In other words, the ``BUY" and ``SELL" decisions can be made more successfully when an accurate movement prediction model is available.
%
\textcolor{black}{For use of $\SMP$ as a trading engine over the $\CDATA$, we consider the following two trading approaches:
\begin{itemize}
\item \textit{The 50-50 Trading Approach:} Whenever  the movement prediction value is less than $0.5$  (i.e., the model senses a plunge in the market's future trends) the ``SELL'' signal will be sent to the trader. Alternatively, when the movement prediction value is greater than $0.5$ (i.e., the model predicts a boost in the market price) it will generate a ``BUY'' signal. We refer to this algorithm as the 50-50 approach. The traders have a chance to buy stocks at a lower price before the price goes up and sell them at a higher price.  Based on these signals,  provided by the $\SMP$ framework, the traders have insightful predictions for the next five days. In other words, based on the $\CDATA$, 5-day in advance predictions can be computed allowing decisions to buy a stock for their portfolio (when predictions show that market will face a ``surge") or vise versa.
\item \textit{The 60-40 Trading Approach:} As an alternative approach, we consider a BUY and SELL with a HOLD scenario, by defining new thresholds over the market movement predictions. Based on this new approach, if the movement prediction value is less than $0.4$ we consider a decrease in the market price. For the values greater than $0.6$, the increase in the market price is predicted. For the values between $0.4$ and $0.6$, the HOLD signal will be offered. We refer to this model as the 60-40 approach.
\end{itemize}}
\section{Experiments} \label{sec:exp}
Experimental results and comparisons are presented in this section to evaluate the proposed hybrid $\SMP$ framework for the task of stock movement prediction. As stated previously, the problem at hand is a classification one with the following expected outputs: (i) On one hand, within a $5$ days prediction horizon, if the adjusted stock price of a specific day is more than that of the previous day, the output of that specific day would be $1$. Then, the sum of the output values is computed over the $5$ days horizon and if the sum is greater than a pre-defined threshold of $3$, we consider the final output for that $5$ day horizon to be $1$, denoting a rise, and; (ii) On the other hand, when the adjusted stock price associated with a specific day is less than its previous day, value $0$ is assigned as the output of that specific day. When the number of such $0$ output values within the $5$ days window is more than $3$, we consider the final output to be $0$, representing the fall prediction/state.

\begin{table*}
\caption {\textcolor{black}{Accuracy, Sensitivity, and Specificity comparisons between the proposed $\SMP$ framework, those of thee stand-alone models, and accuracies of algorithms from References~\citep{Hu:2019, Xu:2018}.}} \label{nnTab:1}
\centering
\begin{tabular}{|l|l|l|l|}
\hline
\textbf{Model } & \textbf{Sensitivity} & \textbf{Specificity} & \textbf{Accuracy} \\ \hline
The $\SMP$ Framework & \textbf{60} & 71.11 & \textbf{66.48} \\ \hline
Standalone CNN Local/Global Model & 56 & \textbf{72.5} & 64.65 \\ \hline
Standalone CNN-LSTM Model & 54.35 & 66.67 & 62.06\\ \hline
Han \cite{Hu:2019} & - & - & 47.8 \\ \hline
Stocknet \cite{Xu:2018} & - & - & 58.23\\ \hline
\end{tabular}
\end{table*}
\subsection{ $\SMP$ LSTM-based hybrid attention Model} \label{sec:EX}
To perform the evaluations, the available Twitter news corpora is tokenized and words occurring less than $5$ times are removed to construct the vocabulary. It is worth noting that removing words with limited usage will reduce the associated memory cost of the DNN models. As stated above, we consider a five day horizon and used  a batch size of  $64$ within $15$ epoch. In addition, Glove, which is an unsupervised word embedding algorithm, is used within the embedding modules of the two parallel paths of the $\SMP$. For comparison purposes, three different models are implemented as follows:
\begin{itemize}
\item[(ii)] \textit{The proposed $\SMP$ Framework}: The proposed hybrid $\SMP$ framework developed in Section~\ref{sec:FingPrnt} is the first implemented stock movement prediction model. The $\SMP$ consists of 2 parallel paths and a fusion path integrating extracted features of each of the two parallel paths.
\item[(ii)] \textit{Stand-Alone CNN Local/Global Model}: The second implemented movement prediction model is the CNN Local/Global path implemented independently (stand alone as a single model). To implement the stand alone version of the CNN Global/Local model, initialization is performed following the guideline provided in Reference~\citep{Seo:2017}. A pre-trained Glove~\citep{Pennington:2014} is used for weighting corpus within the word embedding layer. In the LAL, we use window of size $5$ with a sigmoid function ($\sigma$). Total of $80$ filters are implemented within the LAL. In the GAL, we used $50$ filters of length $2$ and $3$. Finally, a fully connected layer with $0.5$ dropout is designed to form the output.
\item[(iii)] \textit{Stand-Alone CNN-BLSTM Model}: The third implemented movement prediction model is the  CNN-BLSTM path implemented independently as a single model. Similar to the stand-alone CNN Global/Local model, a pre-trained Glove~\citep{Pennington:2014} is used for weighting corpus within the embedding layer. A convolutional layer with a $64$ number of filters and a window size of $25$  is followed by an attention layer. To extract essential features and reduce the framework's complexity, a max-pooling layer is designed. The output of max-pooling layer is the input of next layer which is attention-based Bidirectional LSTM with $250$ hidden layers.  Finally, two fully connected layers are considered with  $300$ and $1$ number of hidden neurons, respectively, to form the price movement prediction results.
\end{itemize}
These three implemented models are trained with Adam optimizer~\citep{Kingma:2014} with a learning rate of $0.001$. To reduce the training times of the implemented models, Batch Normalization~\citep{Ioffe:2015} is utilized to normalize the underlying layers. Furthermore, to avoid overfitting issues and improve the overall robustness of the implementations, $0.5$ dropout is used within the fully connected layers. Finally, the computational graphs of the implemented models are constructed via  Tensorflow~\citep{Abadi:2016} to fine tune different hyper-parameters.
\begin{figure}[t!]
\centering
\mbox{\hspace{-.2in}\subfigure[]{\includegraphics[scale=0.42]{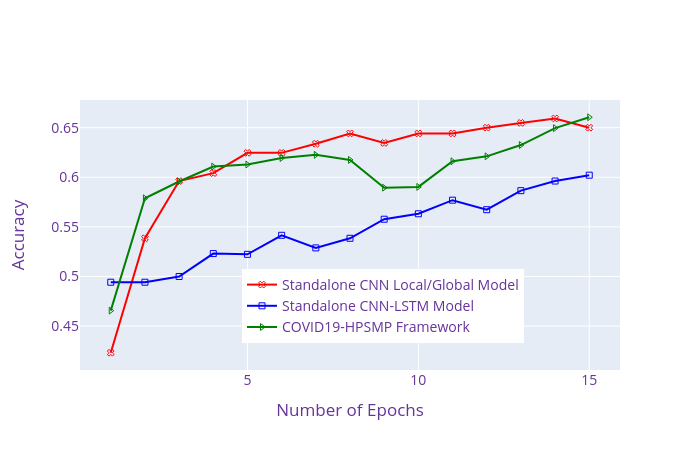}}\hspace{-.6in}
\subfigure[]{\includegraphics[scale=0.42]{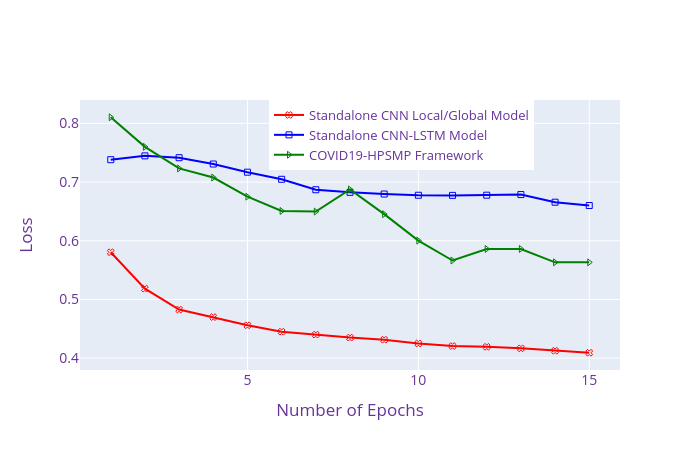}}}
\caption{(a) Accuracy of the Models. (b) Loss of the Models.}
\label{fig:AN8}
\end{figure}

\begin{figure}[t!]
\centering
\includegraphics[scale=0.6]{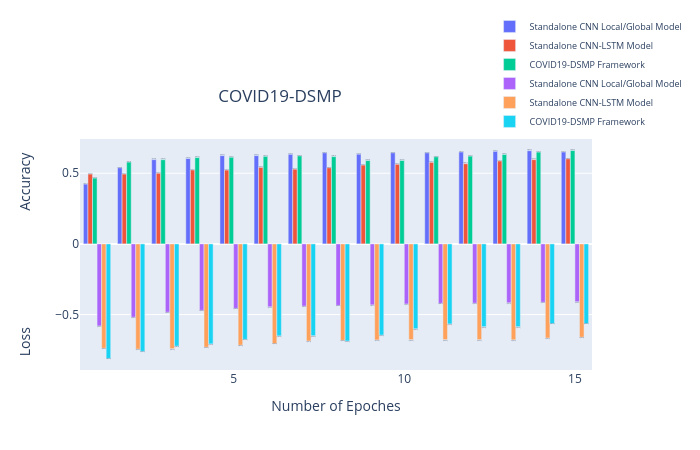}
\vspace{-.4in}
\caption{Accuracy and loss contrast of the different price movement prediction models. \label{fig:Loss}}
\end{figure}
\subsection{Performance Evaluation/Results}
\begin{figure*}[t!]
\centering
\mbox{\subfigure[]{\includegraphics[scale=0.4]{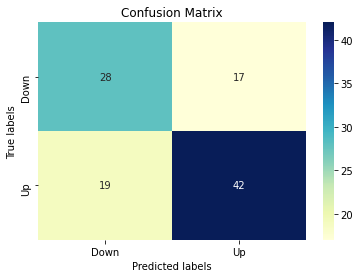}}
\subfigure[]{\includegraphics[scale=0.4]{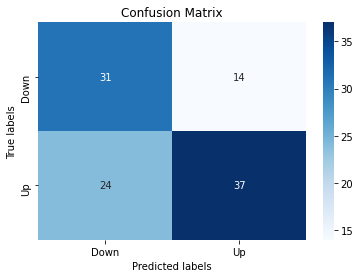}}
\subfigure[]{\includegraphics[scale=0.4]{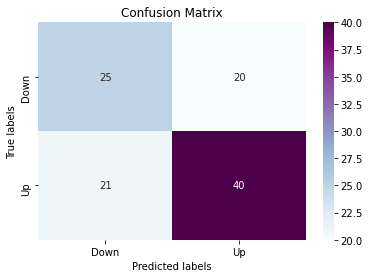}}}
\vspace{-.2in}
\caption{Confusion matrices for binary predictions. (a) The $\SMP$ Framework. (b) Standalone CNN Local/Global Model, (c) Standalone CNN-LSTM Mode.}\label{nFig:CM2}
\vspace{-.25in}
\end{figure*}
In this sub-section, we represent different experimental results to evaluate the performance of the proposed $\SMP$ framework for the stock movement prediction. The accuracy of the proposed models areas follows: $64.65\%$ for the stand-alone CNN-based local/global; $62.06\%$ for the stand-alone CNN-LSTM, and $66.48\%$ for the  hybrid attention model, i.e., the $\SMP$ framework. The accuracy of all three implemented models are shown in Fig.~\ref{fig:AN8}(a). The accuracy is a fraction of correct predictions to the total number of predictions. The loss function associated with the evaluated models is illustrated Fig.~\ref{fig:AN8}(b). Loss function demonstrates the distinction between the output of the model and the target value in order to show the probability of misclassification. We demonstrate the accuracy of the baseline models in Table~\ref{nnTab:1} comparing performance of the three implemented models.

\textcolor{black}{In addition to the accuracy measure, we have computed sensitivity and specificity metrics  to evaluate different error types (i.e., False Negative (FN), and False Positive (FP)).  Sensitivity represents the number of days that stock price has an upward trend correctly predicted by the model. Specificity, on the other hand, is illustrating model's performance in identifying negative conditions. More specifically, if the stock movement test indicates that the price movement has an upward trend while the trend is downward in the actual scenario, the test result is considered to be a FP. Likewise, if the test result indicates a descending trend, but the actual market price is increasing, the result is FN. Sensitivity and specificity are then defined  as follows
\begin{eqnarray}
Sensitivity &=& TP/(TP + FN)\\  \nonumber
&=& (\text{\# of True GoUp Assessments})/(\text{\# of all GoUp Assessments})
\end{eqnarray}
\begin{eqnarray}
Specificity &=& TN/(TN + FP)\\  \nonumber
&=& (\text{\# of True GoDown Assessments})/(\text{\# of all GoDown Assessments}),
\end{eqnarray}
where True Positive (TP) happens when a stock price goes up and the model also indicates the uptrend. On the other hand, True Negative (TN) refers to the scenario where the market indicates that stock goes down and the model shows the downward movement as well. The result of our proposed models are shown in the revised Table~\ref{nnTab:1}. In particular,  we should point out that the higher the numerical value of sensitivity, the less likely the stock movement model returns FP results. The results show that the proposed hybrid model provides considerable improvement in terms of sensitivity, compared to its stand-alone counterparts.}

Fig.~\ref{nFig:CM2} illustrates confusion matrices~\citep{Bhatt:2021} associated to the proposed $\SMP$ framework and its two stand-alone components (implemented individually). As it can be observed, the hybrid model (the $\SMP$) outperforms its counterparts. It is worth mentioning that the achieved accuracy of $66.48\%$ is significant, although in absolute terms it seems to be low. First, please note that average accuracies achieved in the literature for the task of price movement prediction is around $50\%$. Second, these lower accuracies are obtained based on a much wider window of information compared to the limited duration of the introduced $\CDATA$ dataset. The limited duration of the dataset is due to recent emergence of the COVID-19 pandemic.

\vspace{.05in}
\noindent
\begin{figure}[t!]
\centering
\mbox{\subfigure[]{\includegraphics[scale=0.55]{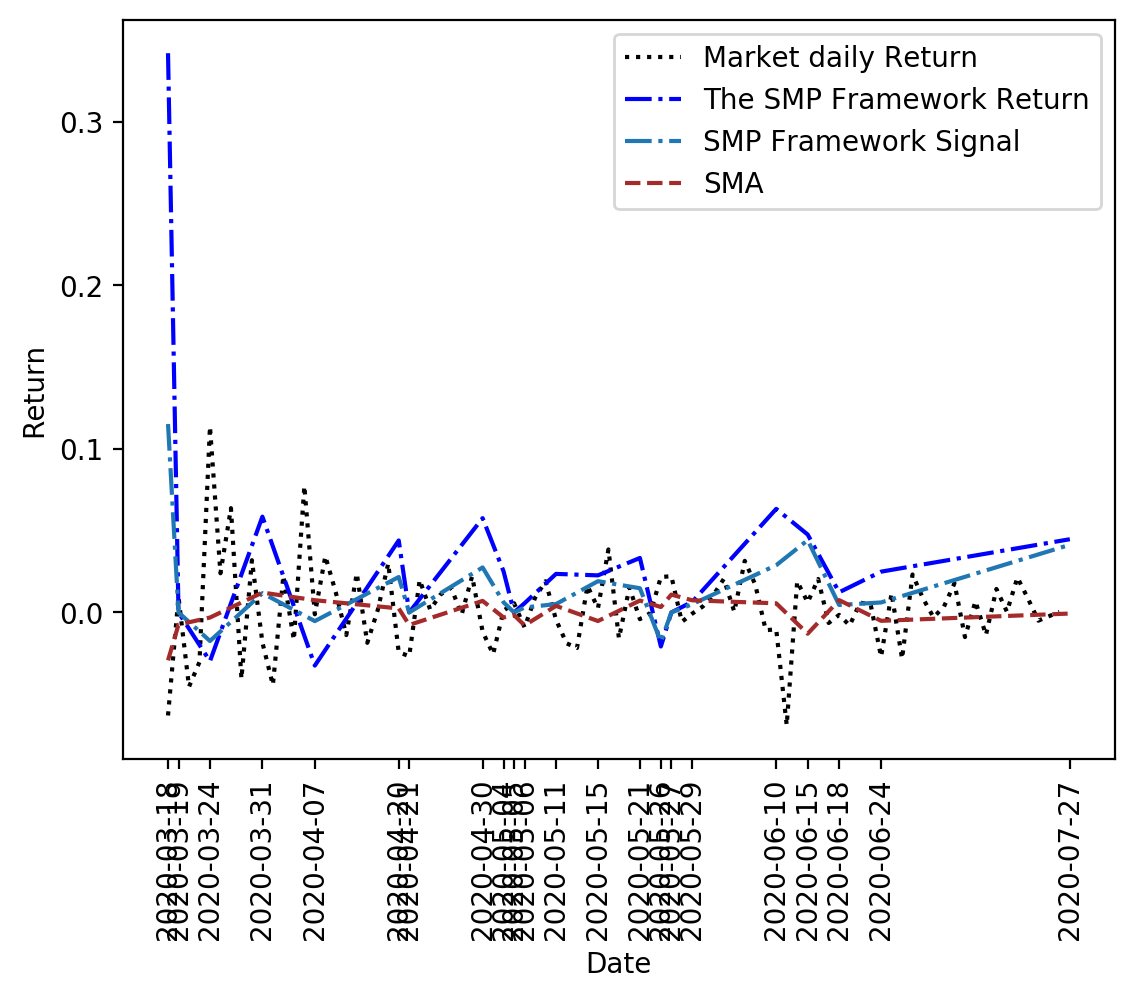}}
\subfigure[]{\includegraphics[scale=0.55]{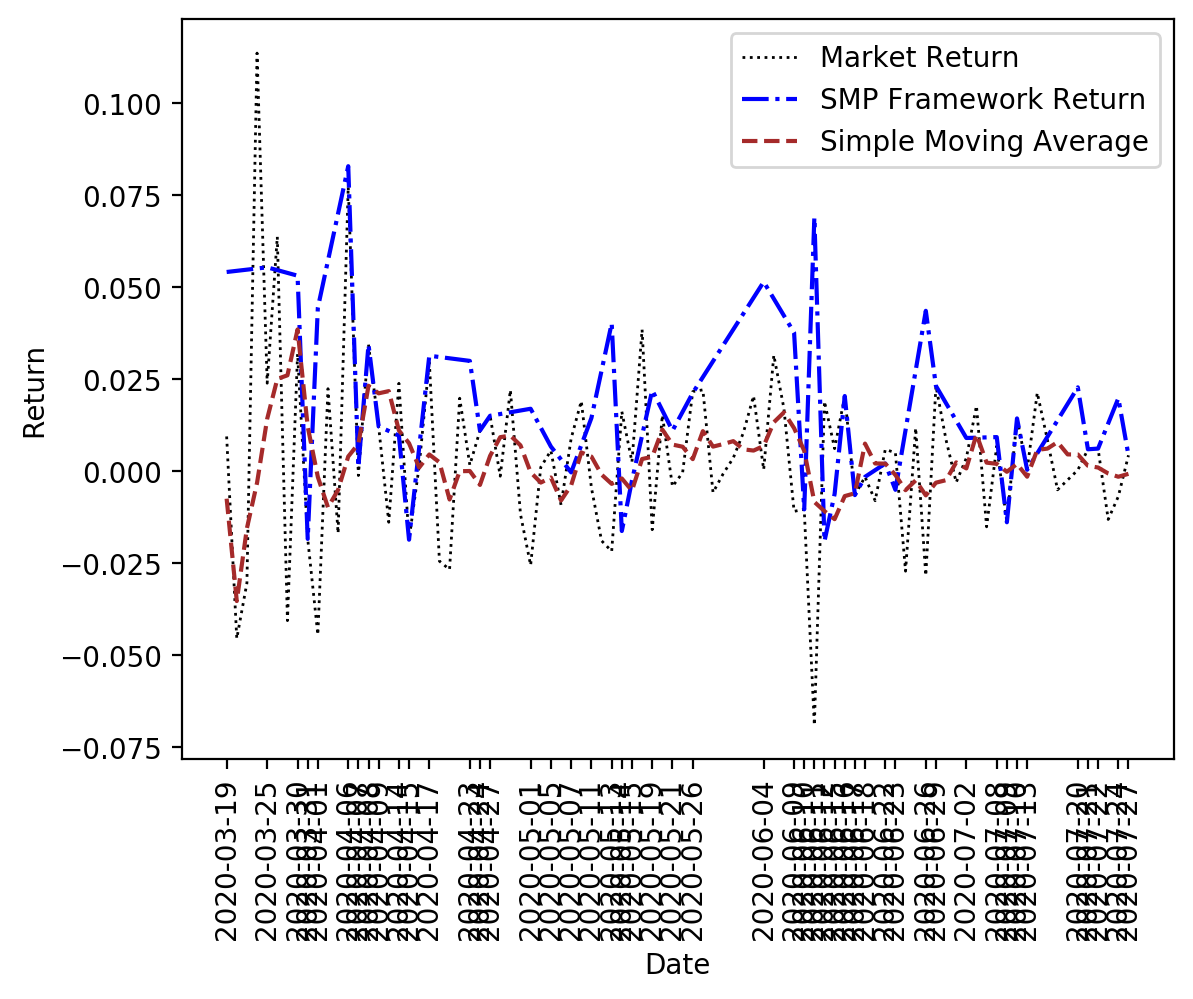}}}
\vspace{-.3in}
\caption{\footnotesize \textcolor{black}{(a) Buy/Sale Return based on the 50-50 approach, i.e., whenever  the movement prediction value is less than $0.5$  the ``SELL'' signal will be sent to the trader. Alternatively, when the model predicts a boost in the market price (movement prediction value is greater than $0.5$) ``BUY'' signal will be generated. (b) Buy/Sale Return based on 60-40 approach, i.e., when the movement prediction value is less than $0.4$ we consider a decrease in the market price sending the  ``SELL'' signal. When the prediction values are greater than $0.6$, the ``BUY'' signal will be sent. For the values between $0.4$ and $0.6$, the HOLD signal will be offered.}}\label{nFig1}\label{nFig6}
\end{figure}
\begin{figure}[t!]
\centering
\mbox{\subfigure[]{\includegraphics[scale=0.6]{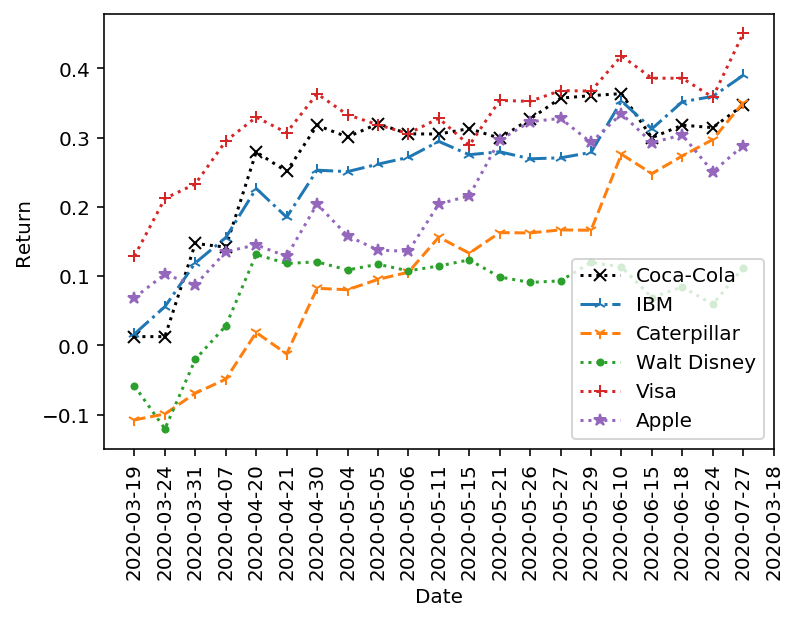}}
\subfigure[]{\includegraphics[scale=0.6]{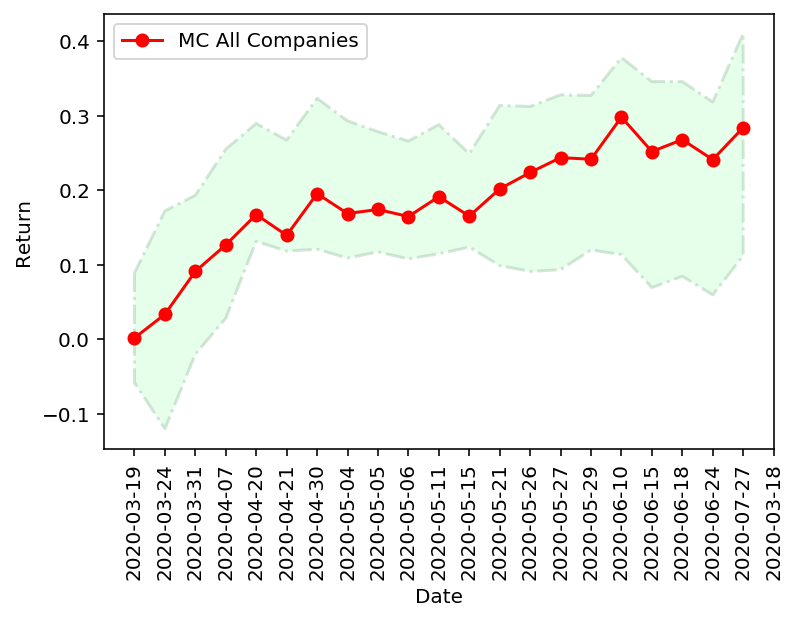}}}
\vspace{-.3in}
\caption{(a) Per-stock returns for the first trading mechanism over the following six prominent companies: $Apple$, $Walt Desney$, $IBM$, $Visa$, $Caterpillar$, and $Coca-Cola$. (b) Results of MC simulations over all companies in Dow Jones.}\label{nFig2}\label{nFig3}
\end{figure}

\begin{table*}
\caption {\footnotesize \textcolor{black}{Experiment results for trading based on the $\SMP$ framework following the 50-50  approach.}} \label{nnTab:2}
\centering
\begin{tabular}{|l|l|l|l|l|l|}
\hline
\multicolumn{6}{|c|}{The $\SMP$ Framework} \\
 \hline
\textbf{Companies} & \textbf{Profit} & \textbf{Return}  & \textbf{Weight} & \textbf{E(R)} & \textbf{Std(R)} \\ \hline
Coca-Cola &   8,009 & 44\% & 8\%  & 3.5\% & 11.5\% \\ \hline
Caterpillar  &  3,790  & 25\% & 18\%  & 4.5\% & 8.7\% \\ \hline
Visa  &  10,606 & 41\% & 28\% & 11.5 & 15.6 \\ \hline
Int. Business Machines Corp. &  7,291 & 39\% & 18\% & 7\% & 13.6\% \\ \hline
Walt Disney  &   3,712 & 19\% & 17\% & 3\% & 6.2\% \\ \hline
Apple & 4,804 & 36\% & 11\% &  4\%  & 10.6\% \\ \hline
\multicolumn{6}{|c|}{Standalone CNN Local/Global Model} \\
 \hline
Coca-Cola & 1,954 & 20\%  & 8\%  & 1.6\% & 5.2\% \\ \hline
Caterpillar  &  566  & 26\% & 18\%  & 4.75\% & 9.2\% \\ \hline
Visa  &2,269 & 24\% & 28\% & 6.7\% & 9\% \\ \hline
Int. Business Machines Corp. & 1,192  & 22\% & 18\% & 3.9\% & 7.6\% \\ \hline
Walt Disney  & 483 & 6\% & 17\% & 1.02\% & 2.2\% \\ \hline
Apple & 1,531 & 16\% & 11\% & 1.8\%  & 4.7\% \\ \hline
\multicolumn{6}{|c|}{Standalone CNN-LSTM Model} \\
\hline
Coca-Cola & 1,651 & 17\% & 8\%  & 1.4\% & 4.4\% \\ \hline
Caterpillar  & 478 & 22\% & 18\%  & 4.1\% & 7.3\% \\ \hline
Visa  & 1,917 & 20\% & 28\% & 5.6\% & 8.6\% \\ \hline
Int. Business Machines Corp. & 1,007 & 19\% & 18\% & 3.4\% & 6\% \\ \hline
Walt Disney  &  408 & 5\% & 17\% & 9\% & 1.8\% \\ \hline
Apple & 1,293 & 14\% & 11\% & 1.5\% & 4.1\% \\ \hline
\end{tabular}
\end{table*}
\begin{table*}
\caption {\footnotesize \textcolor{black}{Experiment results for trading based on the $\SMP$ framework following the 60-40  approach.}} \label{nnTab:3}
\centering
\begin{tabular}{|l|l|l|l|l|l|}
\hline
\multicolumn{6}{|c|}{The $\SMP$ Framework} \\
 \hline
\textbf{Companies} & \textbf{Profit} & \textbf{Return}  & \textbf{Weight} & \textbf{E(R)} & \textbf{Std(R)} \\ \hline
Coca-Cola &   10,448 & 46\% & 8\%  & 3.7\% & 4.9\% \\ \hline
Caterpillar  &  6,896  & 41\% & 18\%  & 7.3\% & 10.2\% \\ \hline
Visa  &  12,590 & 62\% & 28\% & 17.3\% & 23.6\% \\ \hline
Int. Business Machines Corp. &  8,742 & 43\% & 18\% & 8\% & 14.8\% \\ \hline
Walt Disney  &   3,549 & 19\% & 17\% & 3.2\% & 3.5\% \\ \hline
Apple & 7,779 & 59\% & 11\% & 6.5\% & 10.4\% \\ \hline
\multicolumn{6}{|c|}{Standalone CNN Local/Global Model} \\
 \hline
Coca-Cola & 4,978 & 37\%  & 8\%  & 2.9\% & 9.6\% \\ \hline
Caterpillar  &  720  & 28\% & 18\%  & 5\% & 9.8\% \\ \hline
Visa  &3,711& 26\% & 28\% & 7.3\% & 10\% \\ \hline
Int. Business Machines Corp. & 2,304  & 32\% & 18\% & 5.7\% & 7.1\% \\ \hline
Walt Disney  & 1,414 &  9\% & 17\% & 1.5\% & 1\% \\ \hline
Apple & 3,236 & 0.24 & 11\% & 2.6\% & 7.1\% \\ \hline
\multicolumn{6}{|c|}{Standalone CNN-LSTM Model} \\
\hline
Coca-Cola & 3,931 & 21\% & 8\%  & 1.7\% & 5.4\% \\ \hline
Caterpillar  & 764 & 25\% & 18\%  & 4.5\% & 11.7\% \\ \hline
Visa  & 2,027 & 26\% & 28\% & 7.2\% & 9.9\% \\ \hline
Int. Business Machines Corp. & 2,812 & 23\% & 18\% & 4.1\% & 8.1\% \\ \hline
Walt Disney  &  880 & 8\% & 17\% & 1.4\% & 2.7\% \\ \hline
Apple & 2,472 & 17\% & 11\% & 1.9\% & 5\% \\ \hline
\end{tabular}
\end{table*}

\noindent
\textbf{Stock Trading Experiments}: In this part, to evaluate effects of the classification prediction mechanism on the investors' return, we conduct a back-testing experiment by simulating stock trading for March 2020 to the end of July 2020, which is during the COVID-19 financial crisis period. Investors' return~\citep{Zhang:2020} is computed as follows
\begin{eqnarray}
R_{t} &=& w_{t-1} \cdot r_{t}\\
\text{with } \qquad r_{t} &=&\frac{p_{t}}{p_{t-1}}-1, \quad \text{and} \quad w_{t} \epsilon \left[0,1\right]
\end{eqnarray}
where $R_{t}$ denotes the final return of the day, and $r_{t}$ is the return of the day based on the adjusted close price. In the proposed $\SMP$ framework, term $w_{t}$ represents the baseline prediction of the market's movement.  For evaluation purposes, we consider the following three trading  mechanisms:
\begin{itemize}
\item \textbf{\textit{First Trading Mechanism}}: We consider a scenario with $100$ stock shares of a specific company on 01/03/2020. In this regard, first, we perform a selective test over six prominent companies from different sectors, and then conduct a Monte Carlo (MC) simulation. In both cases and to better simulate a real-world trading scenario, we consider a transaction cost of $0.3$\% for each trading.
\textcolor{black}{We have calculated risk-adjusted return based on the Sharpe Ratio (SR) approach. In summary,  the final profit and SR of our $\SMP$ engine is $6,369$ and $1.07\%$, respectively, for the 50-50 approach. The final profit and SR for the second trading mechanism (60-40) is $8,813$ and $2.49\%$, respectively. Generally speaking, SRs greater than one are  preferred with higher the SR value, the better the risk to return scenario for the investors.}
Furthermore, we consider that the stock shares will be available to buy and sell at the moment they are requested. For the first scenario, the following six prominent companies  are selected: $Apple$, $Walt Desney$, $IBM$, $Visa$, $Caterpillar$, and $Coca-Cola$.
\textcolor{black}{Fig.~\ref{nFig1}(a) shows the return obtained based on the $\SMP$ framework and the 50-50 trading approach, compared to the market return.  Fig.~\ref{nFig1}(b) illustrates the results of this 60-40 trading algorithm. As can be seen, the returns obtained based on the 60-40 approach  are higher than the return values gained from the market, standalone CNN Local/Global Model, and standalone CNN-LSTM Model.  The average market return for the 60-40 approach is $0.497$ which shows improvements over the 50-50 algorithm.}
 In Fig.~\ref{nFig1},  the ``BUY" and ``SELL" signals are also shown with green and red triangles, respectively. As can be seen, the $\SMP$ framework closely follows the historical data showing the real market return.  Furthermore, the $\SMP$ engine, in most scenarios, correctly predicts the trend of the market in the future and produces timely ``BUY" and ``SELL" signals. We should point out that the profit is calculated as the difference between the Adjusted Close price of Sell and Buy signals after deducting the transaction cost and risk-adjusted return.
 \textcolor{black}{Further details on the per-stock returns are provided in Table~\ref{nnTab:1} for the 50-50 trading approach, and Table~\ref{nnTab:2} for the 60-40 trading approach. In brief, the final profit and Sharpe Ratio (SP) of the proposed $\SMP$ engine based on the 50-50 approach is $6,369$ and $1.07\%$ respectively. The 60-40 trading approach provides total profit of $8,813$ with SR of $2.49\%$, which means that, overall, the proposed model performs considerably well. }
The $\SMP$ model makes the highest profit for Visa Company, which is $10,606$\$ and the lowest profit for Walt Disney Company is $3,712$\$. Based on the results, we construct a portfolio to measure the overall profitability of the six selected stocks. It is assumed that $100$ shares of each stock are bought, and the trade is made based on the proposed $\SMP$ buy/sell signals. The final profit of our $\SMP$ engine is $6,369$\$, which means that, overall, our model performs considerably well.  In a second scenario, MC  simulation of $100$ runs is performed considering the profit of all the companies in the Dow Jones. In each MC run, six stocks are randomly selected and the procedure outlined above for the first scenario is applied.  As can be seen in Fig.~\ref{nFig3}(b), based on the MC results, the proposed method offers a positive return over this COVID-19 crisis period and can provide the users with a profit despite being in a pandemic area.\\
%
\textcolor{black}{With regards to Figs.~\ref{nFig1} and~\ref{nFig2}, we would like to mention that the market return is computed based on daily trading while the return calculated by the $\SMP$ Framework is based on Buy/Sell signals that the proposed model has produced. In other words, the results shown in the aforementioned figures include the market daily return ($r_{t} =\frac{p_{t}}{p_{t-1}}-1$), which corresponds to using a daily buy/sell signal.  While at times daily trading results in higher return, the proposed model achieved average return of $0.453$, which is significantly higher than the average daily market return of $0.004$.  It is interesting to observe that the $\SMP$ Framework performs better as the model sees more market data. For instance, due to the COVID-19 crisis, abnormal market fluctuations started at the middle of March 2020 and continued until April 2020. Some false predictions in those periods are due to these fluctuations since our model has not seen these type of market reactions during its training phase. After learning from these initial abnormal market fluctuations, the model gradually performs better in predicting the market trend. For example, it can be observed that there is a sharp drop in daily returns in mid of June 2020 where the proposed model's adaptive nature (actively learning) has resulted in achieving positive reruns during that period.}
%
\item \textbf{\textit{Second Trading Mechanism}}: In the first trading mechanism, we considered an equal weight for all the stocks in our portfolio. Here, we construct a portfolio based on market price weighting~\citep{Plyakha:2017} as follows
\begin{eqnarray}
w_i^p &=& \frac{p_i}{\sum_{i=1}^N p_i},
\end{eqnarray}
where $w_i^p$ represents the weight of each stock based on its market price, and $p_i$ denotes the market price (we assume that market prices are adjusted close price of the stock on the first day of March). Weights are applied based on each stock's market price on the date that the transaction is going to be placed. Further details on market price weighting is provided in Table~\ref{nnTab:1}. To evaluate the performance of our model Expected return ($E(R)$) and standard deviation of the return (Std(R)) are computed. The results show that Visa has a profit of $10,606$\$ with standard deviation of $0.156$, which means that visa stock is riskier than others. While the Walt Disney has lower profit and lower risk than the other stocks. Overall, the sum of the profit of the portfolio is equal to $6,764$\$, expected return $E(R)$ is $0.337$ with standard deviation $Std(R)$ of $0.079$.
\item \textbf{\textit{Third Trading Mechanism}}: We consider the scenario where $100$ shares in the Dow Jones market are bought on 01/03/2020 and held until the end of July. Consequently, the cumulative daily return $R$ shows that we get $0.086$ return based on the $\SMP$ framework.
\textcolor{black}{From a finance perspective, the risk also needs to be presented as a model can lead to more returns but higher risk. The risk may be defined as the variability of returns that are expected from a given investment. The greater the variability, the riskier the stock. Investment in the Treasury bond is risk-free because of the governmental guarantees. In contrast, investment in equity shares is risky because of the uncertainty on return. Some shares are riskier than others, and even in years when the overall stock market goes up, many individual shares decline in their price. Therefore, investing in one share is extremely risky. Investment risk, then, is related to the probability of actually earning a low or negative return. Therefore, the probability of low earning or negative return is higher in riskier investments. Consequently, it is necessary to measure the degree of risk associated with the investment. For this purpose, we consider the scenario where $100$ shares in the Dow Jones market are bought on 01/03/2020 and held until the end of July. As a measure of risk, we calculated Sharpe Ratio (SR). The cumulative return and sharpe ratio of our baseline models based on two the trading approaches are illustrated in Tables~\ref{nnTab:4} and~\ref{nnTab:5}. The proposed $\SMP$ based on the 60-40 trading approach obtains the highest cumulative return and SR compared to its counterparts. It is worth mentioning that a higher SR shows superiority of a portfolio compared to its counterparts.}

\begin{table*}[t!!]
\caption {\footnotesize \textcolor{black}{Return comparisons between the proposed $\SMP$ framework and stand-alone models.}} \label{nnTab:6}
\centering
\begin{tabular}{|l|l|l|l|}
\hline
\multicolumn{1}{|c|}{\multirow{2}{*}{Model}} & \multicolumn{3}{c|}{Return}          \\ \cline{2-4}
\multicolumn{1}{|c|}{}                       & 60-40 Strategy & Hold Strategy 1 & Hold Strategy 2 \\ \hline
The $\SMP$ Framework                         & 49.7\%      & 27.1\%      & 8.6\%      \\ \hline
Standalone CNN Local/Global Model            & 41.8\%      & 21\%      & 2.8\%      \\ \hline
Standalone CNN-LSTM Model           & 27.3\%      & 25.5\%      & 2.6\%      \\ \hline
\end{tabular}
\end{table*}
%
\textcolor{black}{As a final experiment, we evaluate the performance  based on the following three trading strategies: (i) \textit{60-40 Strategy:} First, we consider the performance of our models where $100$ shares in the Dow Jones market are bought on 01/03/2020 based on buy, sell and hold scenarios (the so called 60-40 approach); (ii) \textit{Hold Strategy 1:} Second, we consider the ``BUY" and ``HOLD" strategy where $100$ shares in the Dow Jones market are bought and held for a long period regardless of the fluctuations in the market. For this purpose, we bought the stock on the day that the $\SMP$ framework predicted a ``BUY" signal and ``HOLD" it until the end of the period. Based on the signal generated by our model, we buy the stock on 2020-03-19, and sell it on 2020-07-29 when the models predicted a ``SELL" signal, and; (iii) \textit{Hold Strategy 2:} Finally, we consider the scenario where $100$ shares in the Dow Jones market are bought on 01/03/2020 and held until the end of July. The results are shown in Table~\ref{nnTab:6}  illustrating that the proposed $\SMP$ framework achieves higher return across different strategies compared to its counterparts.}

\begin{table*}[t!]
\caption {\footnotesize \textcolor{black}{Return comparisons between the proposed $\SMP$ framework (50-50), stand-alone models and two commonly used trading indicators.}} \label{nnTab:4}
\centering
\begin{tabular}{|l|l|l|}
\hline
\textbf{Model Return} & \textbf{Cumulative Return} & \textbf{Sharpe Ratio} \\ \hline
The $\SMP$ Framework  &   8.6\% & 1.85\% \\ \hline
Standalone CNN Local/Global Model     & 2.8\% & 0.48\% \\ \hline
Standalone CNN-LSTM Model & 2.6\% & 0.41\% \\ \hline
Simple Moving Average (SMA) & 0.9\% &   0.1\% \\ \hline
Moving average convergence divergence (MACD) & 3.8\% & 0.55\% \\ \hline
\end{tabular}
\end{table*}
\begin{table*}[t!]
\caption {\footnotesize \textcolor{black}{Return comparisons between the proposed $\SMP$ framework (60-40), stand-alone models and two commonly used trading indicators.}} \label{nnTab:5}
\centering
\begin{tabular}{|l|l|l|}
\hline
\textbf{Model Return} & \textbf{Cumulative Return} & \textbf{Sharpe Ratio} \\ \hline
The $\SMP$ Framework  &   29.8\% & 2.19\% \\ \hline
Standalone CNN Local/Global Model     & 17.5\% & 0.91\% \\ \hline
Standalone CNN-LSTM Model & 14.9\% & 0.98\% \\ \hline
Simple Moving Average (SMA) & 0.9\% &   0.1\% \\ \hline
Moving average convergence divergence (MACD) & 3.8\% & 0.55\% \\ \hline
\end{tabular}
\end{table*}
\end{itemize}

\begin{figure}[t!]
\centering
\includegraphics[scale=0.80]{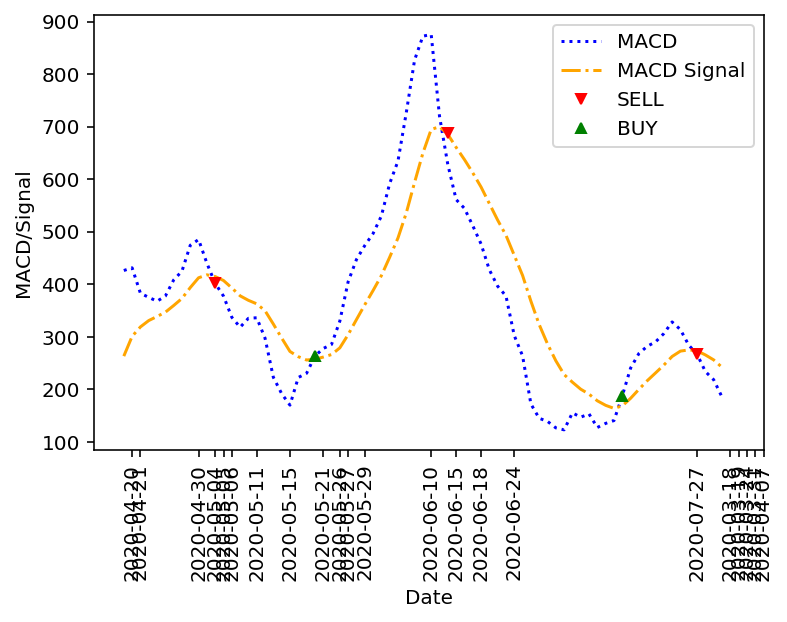}
\caption{MACD Signal}\label{nFig5}
\end{figure}

\vspace{.05in}
\noindent
\textbf{Comparisons with Trading Indicators}: In this part, we compare the $\SMP$ engine with two common trading indicators, i.e., the Simple Moving Average (SMA)~\citep{Ansari:2017} and the Moving Average Convergence/Divergence (MACD)~\citep{Vaidya:2020}, which are among the most widely used technical indicators by market technicians to identify short, medium, and long-term price movements. First we focus on the SAM approach, which is also known as moving mean or rolling mean in statistics. The SMA is one of the oldest and most straightforward strategies for analysing signals and trade based on the results. In this approach, the moving average of the price or return is calculated to estimate the trends in the market price. The SMA is calculated by creating a series of averages of the market data's time-ordered subsets. While different window times can be used, a $5$-day moving average is calculated for comparison with the proposed $\SMP$ framework. Fig.~\ref{nFig6}(b) compares returns obtained from the $\SMP$ framework with those obtained from the SMA approach and that of the market. Fig.~\ref{nFig6}(b)  demonstrates that our $\SMP$ framework outperforms the SMA in prediction of the market's trend.  A second indicator utilized for evaluation purposes is the MACD method, which is another important trading indicator wildly used by traditional technical analysis and algorithmic trading. By leveraging the close price's historical value, MACD can be calculated as a collection of three time-series: (i) The MACD value series, which is the difference between the shorter period (fast) and the longer period (slow) of the exponential moving average (EMA); (ii) The MACD Signal, which is the EMA of the MACD value series, and; (iii) A divergence series, which is the difference between Items (i) and (ii). Three parameters $a, b, c$ are used to calculate the MACD, where $a$ is related to the fast EMA; $b$ relates to  slow EMA, and; $c$ is related to EMA. To compare the proposed framework with the $MACD$ indicator, we used $a = 12$, $b =26$, and $c =9$, which are the commonly used values for these parameters. The BUY/SELL signals generated based on the MACD indicator is shown in Fig.~\ref{nFig5}. The MACD provides five different Buy/Sell signal as shown in Fig.~\ref{nFig5}, which is much less than the Buy/Sell signals provided by the $\SMP$ framework. The cumulative return of different models are illustrated in Tables~\ref{nnTab:4} and~\ref{nnTab:5}. Based on the results shown in Tables~\ref{nnTab:4} and~\ref{nnTab:5}, the $\SMP$ framework outperforms the other two indicators in terms of the cumulative return.

\vspace{.05in}
\noindent
\textbf{Significance Tests}: As a final experiment, we evaluate the hypothesis that predictions made by the $\SMP$ framework are more profitable than those obtained via a standalone CNN Local/Global model (potentially illustrating benefits of a hybrid design), and the Moving Average Convergence/Divergence (MACD) method, a common trading indicator (potentially illustrating benefits in comparison to a commonly utilized indicator). In this regard, we have conducted statistical significance analysis of predictive models given their rate of return through a null hypothesis test. In the first test, the null hypothesis is to use the stand-alone CNN Local/Global engine for investment, and the alternative hypothesis is to use the $\SMP$ framework as a recommendation engine for investment. The price movement prediction in stock market is assumed to be aleatory (null is not rejected) until it is proven the results of a recommendation engine (in this case, the $\SMP$ framework) has no relationship with the results of a random trader (null is rejected). To define a confidence threshold, a cut point of $\alpha$=2.145  corresponding to 95$\%$ of confidence is established. The resulting $t$-value is equal to $3.0526$, which is associated with $p$-value of $0.004303$ showing a significant result as $p < 0.05$. This suggests that the null hypothesis is not accepted, therefore, predictions made by $\SMP$  are more profitable. Similarly, in a second experiment, we check if predictions of the $\SMP$ framework  is more profitable than those of the MACD indicator. The computed $t$-value is $2.8433$ (i.e., $p$-value of $0.00651$), which means that the $\SMP$ framework reaches better results compared to the MACD, which is a common trading indicator.

\textcolor{black}{As a final note, we would like to mention that the variance between the actual return from a given portfolio and the expected return over a given period of time is referred to as an ``Abnormal Return''. A positive abnormal return refers to the scenario that the actual return is higher than its expected return. Abnormal return can be achieved in stock market, however, the volatile nature of the  market is hard to earn abnormal return using only the financial indicators, i.e., the stock data, as the market is dependent on a variety of social, political, and economical factors. There are three forms of efficient markets (i.e., weak, semi-strong, and strong), defined based on the nature of information utilized to determine asset prices. In the weak form, asset prices referring to all past prices and trading volume information fully reflect all the market data. In the semi-strong form, asset prices reflect all publicly known and available information. In the strong form, asset prices fully reflect all information, including public (e.g., news) and private information. We consider semi-strong form of efficiency for the markets, in which prices reflect the news data and financial indicators. The proposed $\SMP$ framework incorporates effects of social media trends related to COVID-19 on stock movement prediction. The proposed framework uses information fusion to combine COVID-19 related Twitter data with historical data to trade 5-day in-advance in order to predict the movement of the market. Our model can predict the stock market's fluctuations with more than 66\% accuracy, which will hopefully be a metric to earn abnormal return.}
\section{Conclusion} \label{sec:conc}
Motivated by abrupt, sudden, and negative effects of COVID-19 pandemic on stock markets, first, the paper introduced a unique COVID-19 related PRIce MOvement prediction ($\CDATA$) dataset. The constructed dataset incorporates effects of  social media trends related to COVID-19 on stock market price movements. Based on the constructed $\CDATA$ dataset, the paper then proposed a novel data-driven (DNN-based) COVID-19 adopted Hybrid and Parallel deep  fusion framework for Stock price Movement Prediction ($\SMP$). The proposed framework uses information fusion to combine COVID-19 related Twitter data with extended horizon market historical data. More specifically, in contrary to the existing data-driven stock price movement prediction models, where a single DNN model is used, the $\SMP$ framework is a hybrid model consisting of two parallel paths (i.e., the CNN Local/Glocal path, and; the CNN-LSTM path) and a fusion path that combines localized features. Each of the two parallel paths is a unique attention module extracting different attention related features. The rationale behind such a hybrid and parallel structure is the significance of the attention network and the intuition that extracting different attention-related features would improve the overall performance of the model. The proposed $\SMP$ architecture can predict the stock price movements during the pandemic crisis to forecast sudden sharp movements (fall or rise) in the stock market. Based on the results of the $\SMP$ architecture, we can predict the stock market's fluctuations with more than $66\%$  accuracy, which will hopefully be a metric to be more prepared for the unexpected havocs.

\bibliographystyle{model5-names}
\biboptions{authoryear}

\end{document}